\documentclass[a4paper]{jpconf}
\usepackage[utf8]{inputenc}
\usepackage[T1]{fontenc}
\usepackage{lmodern}
\usepackage{graphicx}
\usepackage{amsmath}	
\usepackage{amssymb}	
\begin{document}
\title{The Future of Black Hole Astrophysics in the LIGO-VIRGO-LPF Era}
\author{Roger Blandford}
\address{KIPAC, Stanford, CA 94305}
\ead{rdb3@stanford.edu}
\author{Richard Anantua}
\address{Dept. Astronomy, UC Berkeley, CA 94720}
\ead{ranantua@berkeley.edu}
\begin{abstract}
There is a resurgence of interest in black holes sparked by the LIGO-VIRGO detection of stellar black hole mergers and recent astronomical investigations of jets and accretion disks which probe the spacetime geometry of black holes with masses ranging from a few times the mass of the sun to tens of billions of solar masses. Many of these black holes appear to be spinning rapidly. Some new approaches are described  to studying how accreting black holes function as cosmic machines paying special attention to observations of AGN jets, especially with VLBI and $\gamma$-ray telescopes. It is assumed that these jets are powered by the electromagnetic extraction of the spin energy of their associated black holes, which are described by the Kerr metric, and that they become simpler and more electromagnetically dominated as the event horizon is approached. The major uncertainty in these models is in describing acceleration and transport of relativistic electrons and positrons and simple phenomenological prescriptions are proposed. The application of these ideas to M87 and 3C279 is outlined and the prospects for learning more, especially from the Event Horizon Telescope and the Cerenkov Telescope Array, are discussed. The main benefit of a better understanding of black hole astrophysics to the LISA mission should be a firmer understanding of the source demographics.
\end{abstract}
\section{Introduction}
The last time one of us (RB) attended a LISA meeting was a decade ago and we would like to start by summarizing what was said then just to give give some context to how things have changed in studying both gravitational radiation and black holes. In 2006, there was an expectation that LIGO would make the first direct detection of gravitational radiation, as has happened, but that LISA sources (white dwarf binaries, EMRIs, MBH binaries, stochastic background and, of course, the unanticipated) would be more diverse, could be studied in greater detail and would directly impact more astrophysical questions.  RB highlighted what he called ``harbingers'' -- advanced notice of incipient mergers in distant AGN with few degree positions from LISA waveforms -- and the use of LSST images to seek an electromagnetic signal in phase with the waveform ahead of the merger and thereby make an identification. It now looks like this will not be practical because the anticipated accuracy of the LISA positions has been downgraded. However the search for ``afterglows'' will still be practical and optimism remains high that these will be seen. In this connection RB resurrected an old idea that merging black holes might detonate enormous electromagnetic explosions -- a tiny fraction of the gravitational radiation energy release would suffice -- by acting as two-sphere or three-sphere dynamos. Although he expressed some skepticism, this might allow a seed electromagnetic field to grow exponentially, just prior to the merger, provided that the effective resistive decay did not quench the growth. Recent simulations (East and RB, in preparation) demonstrate that there is no such growth. This result is supported by observation!  RB also discussed resonant excitation of stellar oscillations when white dwarfs follow elliptical orbits around more massive black holes \cite{Rathore2005}. The resonances are inevitably crossed as gravitational radiation shrinks the orbit and the energy transfer could be large enough to explode the white dwarf, especially when the mass ratio is large. We still think this is an interesting possibility and could lead to a new class of supernova explosion.

Observationally, RB described Very Long Baseline Interferometry (VLBI) with its improving resolution and sensitivity. Here the big development has been great strides in mm interferometry.  It has already been found that the M87 jet appears to be collimated on scales less than of order ten gravitational radii and that the Sgr A$^{\ast}$ polarization is high, implying an organized magnetic field close to the black hole\cite{Akiyama2015}\cite{Johnson2015}. We should soon have even more impressive images and model fitting for these and other sources especially under the aegis of the Event Horizon Telescope (EHT). JWST, which then had a scheduled 2013 launch is now expected to start observing in 2019 but may have little overlap with LISA, as currently planned. The wide field OIR space telescope JDEM has become Euclid, to be launched at the start of the next decade and WFIRST, scheduled for mid-decade. The next generation X-ray proposals, Con-X and XEUS have evolved into Athena, which is now seen as being on a similar timescale to LISA, late decade. GLAST was launched in 2008 as Fermi Gamma-ray Space Telescope  (FGST) and opened up the GeV sky. As a widefield instrument, it sets important limits on afterglows but is now unlikely to overlap with LISA. TeV astronomy with Atmospheric and Water Cerenkov detectors was also discussed and there have been great advances with the discovery of many new blazars as well as many intriguing Galactic sources. Here, the future lies with the Cerenkov Telescope Array (CTA), two observatories that will soon produce their first data and which should be fully operational in the early part of the next decade. The optimism for neutrino astronomy has been vindicated, especially with the successful deployment of Ice Cube and its detection of (still unidentified) cosmic sources.  

The bad news, though, is that the launch date of LISA has been pushed out from the (2006) nominal date of 2016 by at least a decade. Others here discussed the prospects for accelerating and funding the project, especially following the great success of LPF and the scientific impetus of GW150914. The era of gravitational wave astronomy, with its associated physics program, has begun and there is a a whole spectrum extending from the LIGO band through the LISA, IPTA and CMB bands to explore. 
\section{Black Hole Astrophysics}
\subsection{Generalities}
Although once regarded with suspicion by some theoretical physics intimidated by the perplexing questions they raised and dismissed by the more conservative observational astronomers, black holes are now part of the astrophysics mainstream. We are pretty sure that we are witnessing the births of stellar holes on a daily basis as $\gamma$-ray bursts and, even more frequently, as a minority of core collapse supernovae. The evidence that normal galaxy nuclei contain massive black holes is just as compelling. The few GW events provide an even stronger existence proof and, more importantly, they confirm that general relativity can be used to describe them. The masses that we observe are in the range $\sim4{\rm M}_\odot-2\times10^{10}{\rm M}_\odot$. Spin measurements are model-dependent and probably less accurate but do suggest that many holes have large spins, $a\gtrsim0.9m,\Omega_H\gtrsim0.3m^{-1}$. Even less secure are the basic principles of disk accretion despite considerable, thought, analysis and simulation. We still do not have a description of the physics at the level of our description of main sequence stars. 

Most instances of black hole accretion are believed to have enough angular momentum to form an extended disk. There appear to be three main accretion modes, mostly distinguished by the mass supply rate in units of the ``critical'' or ``Eddington''  or ``Salpeter'' rate, $4\pi GMm_p/\sigma_Tc^2\equiv\dot m$. (Less important factors include the hole mass and spin, the outer disk radius in units of the gravitational radius $GM/c^2\equiv m$, the composition of the accreting gas and environmental factors, especially in for stellar holes.) In the case of AGN, it is clear that overall disk accretion has to be very efficient - $\sim0.1-0.3c^2\equiv10-30{\rm\ PJ\ kg}^{-1}$, because we observe the radiant energy density and measure the local hole mass density \cite{Soltan1982}. However, there is now pretty strong evidence that accretion disks are only thin down to their inner radius and are capable of radiating their released gravitational binding energy on the inflow timescale when the mass supply is intermediate. $0.01\lesssim\dot m\lesssim1$, say.  At lower mass supply, the ion-electron energy exchange time is longer than the inflow time and a thick accretion torus seems to form. At higher mass supply, radiation is efficient but escape is slow on account of the high Thomson optical depth and the radiation is advected with the flow. So, at high and low $\dot m$, the accretion flow is roughly adiabatic. Under these circumstances, the torque that drives the accretion must transport energy outward and the release of this energy is enough to unbind the gas in the disk\cite{Blandford1999}. There has to be an outflow to carry away the accretion energy when radiation fails to do the job. In the low $\dot m$ case, this makes cooling even harder and the mass accretion rate can be many orders of magnitude smaller than the supply rate. This appears to be the case in Sgr A$^\ast$ and M87 where there is no conclusive observational evidence for an accretion disk close to the hole.  

It is also clear that the internal disk torque is hydromagnetic because weakly magnetized disks are generically subject to the magnetorotational instability which grows the field on a rotational timescale\cite{Balbus1991}. It is also quite likely that this magnetic field is not confined to the disk and, instead  is responsible for an external torque and an outflowing wind. It is difficult to be quantitative. Magnetic flux may also be trapped within the inner disk radius and can thread the event horizon of the black hole. When the hole spins, electromagnetic power is extracted and this appears to be responsible for relativistic jets\cite{Blandford1977}. The bulk Lorentz factors in the jets associated with AGN and of most interest to the LISA community are $\Gamma\sim10$; those associated with X-ray binaries have $\Gamma\sim3$ while$\Gamma\sim300$ for $\gamma$-ray bursts. The jets are thought to be confined by a thick accretion torus and magnetized outflows. In this interpretation, the high latitude magnetosphere at the base of a black hole jet is likely to be electromagnetically dominated as low altitude plasma will drain across the horizon and high altitude plasma will fly away, being replenished by electron-positron pair production or cross-field transport of ionic plasma within the black hole magnetosphere. In addition to the observational evidence that relativistic jets involved strong magnetic field we are starting to see evidence that pair plasma is also present\cite{Siegert2016}.
\subsection{Electromagnetic Jets} 
When a black hole with angular frequency $\Omega_H$ is threaded by magnetic flux $\Phi_H$ sourced by external toroidal current, an EMF $V\sim\Omega_H\Phi_H$ will be generated. Under electromagnetic conditions, a current $I\sim V/Z_0$, where $Z_0$ is the impedence of free space will flow and the extracted power, derived from the hole's rotational energy, is $\sim V^2/Z_0$. A comparable power will be dissipated within the event horizon, increasing the mass of the hole.  AGN jets typically have $\Phi_H\sim10^{25-28}$~Wb and $V\sim1-100{\ \rm EV}, I\sim 10{\rm\ PA} - 1{\rm\ EA}$. Particle acceleration is not a challenge given these large voltages. Indeed the most powerful AGN jets remain a candidate site for the acceleration of Ultra High Energy Cosmic Rays. Another feature of this mechanism is that the angular velocity associated with the field lines (that of a reference frame in which the local electric field vanishes) will be $\sim0.5\Omega_H$, being larger for polar than low latitude field lines. Three (space) dimensional general relativistic MHD simulations bring out and quantify these general features\cite{McKinney2012}. In particular they confirm that fairly close to the hole, within $\lesssim100m$, say, the jet structure is quite simple and electromagnetically-dominant. In addition the jets are surpringly robust. This will quite likely be a boon when the EHT investigates jets closer to their originating black holes. In fact these inner jets can be described approximately by a simple analytical solution with a paraboloidal profile that brings out the main features of the simulations. In addition, it is found the return current from the jet is concentrated close to the jet wall. In other words, the electromagnetic stress within the jet is balanced by the external gas pressure. 

There are three consequences of the presumed electromagnetic dominance close to the event horizon. The first is that the commonly invoked diffusive shock acceleration is irrelevant. Secondly, while the electromagnetic field and current structure may be fairly simple and well-defined, the fluid (plasma) component is quite unknown. There is no prescription for specifying where and how this plasma is created though the amount needed to supply electrical current and provide space charge is tiny. However, we can say something about the velocity of this plasma, presuming that it can be treated as a single fluid.  The component of velocity measured in a local inertial frame perpendicular to the presumably dominant magnetic field will be given by $\vec E\times\vec B/B^2$.  The component along the magnetic field may also be fairly well-defined except close to the black hole. In the perfect MHD limit, it will be essentially the mechanical speed of a bead on a rotating wire with the same shape as the magnetic field. Of course, this depends upon the initial conditions and instabilities, radiation reaction or particle drift may vitiate MHD but it is a good starting assumption that the velocity field, as furnished by a simulation is a fair representation of that found in an actual relativistic jet. 

The third consequence that emerges from these models is that at larger distances from the hole, $\gtrsim1000m$, say, the jets build up internally created pairs, interact strongly with their environment and are subject to instabilities, which, while not completely disrupting them, do lead to much entrainment. We expect the jets to accelerate first then to decelerate. As the jet inertia builds up, more of the jet current closes within the jet, thereby accelerating the plasma. Note that the jet speed is quite variable in space and time in contrast to what is commonly assumed. 
\section{"Observing" Numerical Simulations}
We have argued that RMHD simulations now provide a good foundation for modeling a jet close to the hole where the gas inflow and the electromagnetic outflow might be relatively simple. In order to compare the model with actual observations -- to ``observe'' the simulation -- it is necessary to add a prescription for particle acceleration\cite{Anantua2016}. There are several such prescriptions that can be used. 
\subsection{Particle Acceleration}
Let us consider the synchrotron emission and suppose that the the particle distribution function is isotropic. It is helpful to measure the particle density by the partial pressure $\tilde P(\nu)$ of the electrons and positron with energy such that emit syncrotron radiation in the local magnetic field within an octave of frequency of the actual observed frequency. This focuses attention on the electrons that have to be present and must have been accelerated instead of a much more extensive distribution with density that we cannot measure. A variety of prescriptions have been entertained. The simplest is that $\tilde P(\nu)$ is a simple fraction of the magnetic pressure $P_M$. (The further assumption that $\tilde P(\nu)$ is independent of $\nu$ is essentially equivalewnt to saying that the observed spectral index is 0.5.) This is a natural generalization of the ``equipartition hypothesis''. A further extension is to suppose that $\tilde P(\nu)\propto P_M^n$. 

Other prescriptions are based upon electromagnetic generalizations of fluid dissipation prescriptions. For example, following the practice with accretion disks, we can suppose that the Maxwell shear stress is a fixed fraction -- $\alpha$ -- of $P_M$. The rate of dissipation can then be estimated by the product of this stress and the transverse velocity gradient. Alternatively, we can introduce a coefficient of shear viscosity $\mu$ as some multiple of $cL$ where $L$ is the characteristic length scale associated and use the full Maxwell stress tensor. An electromagnetic prescription introduces a resistivity $\eta\sim\mu_0cL$ and evaluates the rate of dissipation using $\eta j^2$. These estimates of the rate of dissipation can be converted into an estimate of $\tilde P(\nu)$ using an acceleration time scale which is most simply chosen to be the shorter of the expansion time scale and the radiative cooling timescale.

These essentially phenomenological models of particle acceleration lead to quite varied jet behavior when they are ``observed''. It is hoped that with future observations, they will help us understand the microphysical processes at work when particles are energized.
\subsection{Radiative Transfer}
The treatment of radiative transfer should be straightforward but new approaches have been developed to provide an efficient interface with a relativistic hydromagnetic simulation. The equation of polarized radiative must be solved in the frame in which the simulation is performed, essentially that of the host galaxy. However, the emissivity and absorption coefficient are specified in the local rest frame -- the comoving frame --  inside the jet, and have to be transformed allowing for the Doppler shift, aberration and rotation of the plane of polarization. They must be expressed in terms of electromagnetic variables evaluated in the comoving frame.
\subsection{mm Observations}
The accreting, massive black holes in the nuclei of active galaxies create antiparallel jets, most commonly observed at radio frequencies. The smallest scale structure at the base of these jets is resolved using VLBI and it is now possible to use this technique at mm wavelengths with intercontinental baselines thereby exploring the mechanisms at work in powering and collimating them.  We already know that the jets are collimated quite close to the black hole and the jet emission can be limb-brightened. 

Imaging polarimetry at mm wavelengths promises to be very prescriptive. The linear polarization should help us to determine if the magnetic field has the basic structure predicted, becoming increasingly toroidal with increasing cylindrical radius. In particular, as the plasma near the hole is moving slowly or even inwards. we may be able to observe the counterjet\cite{Broderick2016}, gravitationally lensed by the black hole (as well as the torus). This should also allow us to see if the magnetic field geometry is basically dipolar (as generally expected) or quadrupolar. The inferred sense of rotation of the hole is expected to be the same as that of the gas observed to be orbiting the hole at large radius.  If circular polarization can be measured, then it may distinguish an electron-positron plasma from an electron-ion plasma especially if the two frequencies can be observed. 
\subsection{$\gamma$-ray Observations}
The small minority of jets that are directed towards us are called blazars. (However, Doppler beaming allows them to dominate flux-limited samples.) Blazars were originally studied using radio and optical telescopes. However, many of them are most prominent in$\gamma$-rays both at GeV energies, as observed by FGST and at TeV energies from the ground. One of the big surprises was how many of the sources are extremely variable with time scales as short as a few minutes. This suggests to many, though not all, astronomers that the emission comes from close to the black hole. However, it must originate sufficiently far out to avoid pair production on the photon background. Now most blazars have ``Bactrian'' (two humped) spectra when plotted as $\log(\nu F_\nu)$ as a function of $\log(\nu)$ ranging from $\sim8$ to $\sim25$. It is conventional to associate the low frequency hump with synchrotron radiation and the high frequency hump with Compton scattering. It is also generally supposed that the soft photons that are being scattered mostly originate outside the jet, originally as disk photons -- the External Compton representation -- in the higher power ``Flat Spectrum Radio Quasars'' (FSRQ) and that they mostly originate inside the jet -- the Synchrotron Self Compton representation -- in the lower power ``BL Lac'' sources. 
\section{M87}
The M87 jet has been studied carefully for a century. The black hole mass is measured to be $M\sim 6.6\times10^9{\rm M}_\odot\equiv m\sim10^{13}{\rm m}\equiv4\mu{\rm as}$\cite{Gebhardt2011} and is now well-imaged down to projected radii $\sim100m$ and modeled down to $\sim10m$. Fitting a stationary analytical approximation to a numerical simulation suggests $\Omega_H\sim10^{-5}{\rm rad\ s}^{-1}$, $\Phi_H\sim10^{26}{\rm Wb}$, $V\sim50{\rm EV}$, $I\sim1{\rm EA}$. The outflow speeds at jet radii $\sim100m$ are $\sim0.7-0.95{\rm c}$. The M87 jet and, by inference, the hole spin axis is inclined at an angle $\sim20^\circ$ to the line of sight. This means that Doppler boosting and polarization rotation will be important. Nevertheless, if the magnetic field is organized, as we conjecture, then the polarization will be large and we should be able to learn details of its geometry.

One feature of the models that is both striking and robust is the presence of a lateral shift of the ``spine'' of the jets due to the dependence of the single particle emissivity on the pitch angle which has to be different on the two sides of the jet. (See Fig. 1.) The observations should also provide indications of entrainment. The innermost parts of the jet are self-absorbed and so one is really viewing a radio photosphere. as this will move inward with increasing radio frequency, the centroid of the jet ``core'' emission will move closer to the hole as the frequency of observation is increased. 
\begin{figure}[ht]
\begin{center}
\includegraphics[width=30pc]{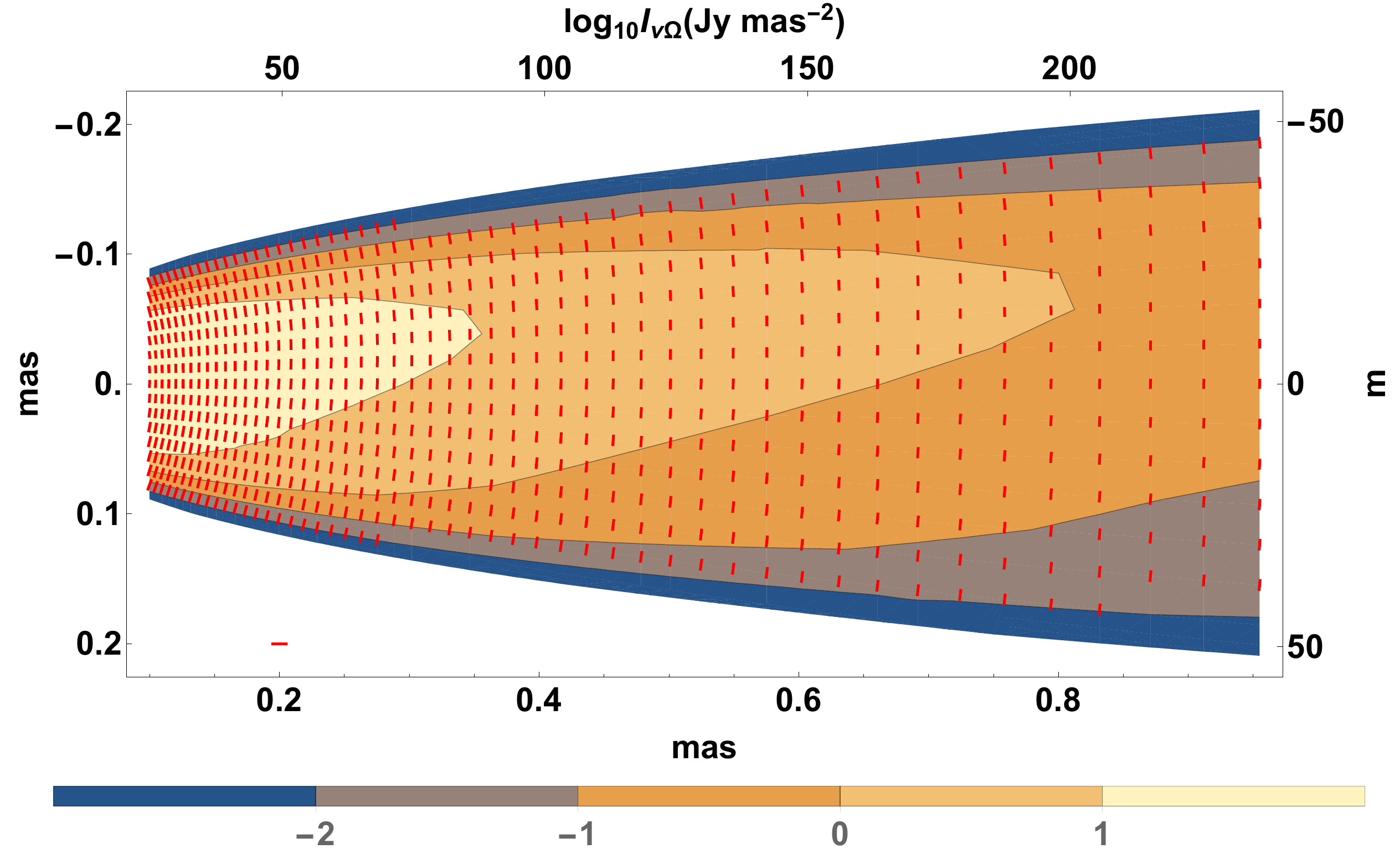}
\end{center}
\caption{\label{lab}Simple model of the M87 jet that shows a vertical shift in the spine of the jet relative to the jet axis due to the dependence of the synchrotron emisssivity on pitch angle. The red bars represent linear polarization, with the isolated bar at the bottom corresponding to 100\% polarization.}
\end{figure}
\section{3C279}
3C279\cite{Ackermann2016} is a good example of a $\gamma$-ray very luminous FSRQ. It has an optical luminosity of $\sim3\times10^{38}{\rm\ W}$ and an isotropic $\gamma$-ray flux that varies between $\sim10^{40}{\rm\ W}$ and $\sim10^{42}{\ \rm W}$ , consistent with strong relativistic beaming. The mass of the black hole is estimated to be $\sim10^9{\rm\ M}_\odot$ and a bulk Lorentz factor $\Gamma\sim30$ is indicated during major flares. In the simplest models all the jet components are produced by a single moving source. Consequently, the magnetic energy density would have to be much lower than the soft photon and the particle energy density. A recent discovery is that occasional $\gamma$-ray flares show variability on time scales (after correcting for cosmological redshift) as short as $t_{\rm var}\sim3{\rm \ min}\equiv m/30$. Relativistic kinematics allows this rapid a variation to arise at a radius $r\sim\Gamma^2ct_{\rm var}$. The conclusion from more detailed calculations is that this can be just compatible with a minimalist model of the soft photon background. 

In order to reconcile these observations with the electromagnetic jet model, it is nexcessary to consider alternative models for the emission. The first possibility to consider is that the ``synchrotron hump'' originates from much larger radii than the $\gamma$-rays. There is no evidence for minute-scale variation at these much lower frequencies. The second modification is to suppose that the $\gamma$-ray emission is synchrotron radiation. There is a precedent for this as the variable $\sim300{\rm\ MeV}$ emission from the Crab Nebula is  almost certainly made by the synchrotron process. However, it is commonly claimed that synchrotron photons have a maximum energy $\sim\alpha_F^{-1}m_ec^2\sim70{\rm\ MeV}$ which may be hard to reconcile with $\sim{\rm\ TeV}$ emission, even after allowing for a Doppler-boost of the photon energy by a factor $\sim\Gamma$. This conclusion follows from noting that the radiation reaction force on an electron emitting photons with characteristic frequency $\cal E$ can be expressed as $\sim({\cal E}\alpha_F/m_ec^2)ecB$. This must be balanced by an accelerating force $\sim eE$ and the expectation that $E\lesssim cB$ delivers the result. 

Before examining this argument more carefully, let us illustrate it with some quite model-dependent numbers. Suppose that the $\sim300{\rm\ MeV}$ flare emission comes from $r\sim100m$. A jet that is still electromagnetic will have a a comoving field strength $\sim3{\rm\ mT}$ and be emitted by electrons with comoving energy $\sim3{\rm\ TeV}$. The comoving radiative cooling time is $\sim0.1{\rm\ s}$. Short, sporadic electric fields with strength $\sim 0.1cB$ are sufficient to accelerate the particles to these energies and dissipate the electromagnetic energy density, a generic process called ``magnetoluminescence''\cite{Blandford2014}.   Specific mechanisms that achieve this can be exhibited using ``Particle in Cell'' simulations\cite{Yuan2016} for example. The most extreme acceleration involves the creation of occasional linear magnetic-field parallel electric field sites. Electrons are rapidly accelerated and radiate when they escape these sites and are scattered into large pitch angle gyrastions about the magnetic field. In our example, freshly accelerated electrons should continue to cool down to energies $\sim20{\rm\ MeV}$, creating a spectrum $\nu F_\nu\propto\nu^{0.5}$ in the simplest case, when expansion losses will be more important than radiative losses.

Powerful blazars like 3C279 emit $\gamma$-rays up to energies $\sim1{\rm\ TeV}$. These are not known to be so variable and can originate from larger jet radii. Can these be synchrotron photons too?  Two  variations on the scheme just sketched can be mentioned. The first is that the electrons are impulsively accelerated by electric fields directed along the comoving magnetic field and so develop small pitch angles alllowing them to achieve higher energy and emit higher energy photons before radiation reaction becomes significant. They can be scattered in pitch angle by plasma instability and then start to radiate very efficiently. The second possibility is that it is protons that are accelerated so that direct radiation reaction is unimportant. They could be accelerated to vey high energy, even EeV. The major loss may arise from electron-positron pair production by soft photons interacting with the proton Coulomb fields. This would initiate electromagnetic showers. In general electromagnetic dissipation at small radius is likely to create showers until the $\gamma$-rays can escape from the jet.  More detailed simulations are clearly needed to determine if models like these are viable\cite{Anantua2017}, for example. 
\section{Summary and Outlook}
Although, in the popular imagination, black holes are associated with closure, astronomers regard them as vital entities. From their proud birth announcements as $\gamma$-ray bursts and GW events, through the brilliant lives that some of them undergo to the unlikely, though not impossible, new contributions that they could yet make to fundamental physics as primordial, cosmological agents, firewalls and even macroscopic wormholes, there is a renaissance in black hole astrophysics. The energy released by the accreting black hole in the nucleus of a typical galaxy is comparable with that released by the stars (and ten times the gravitational binding energy of the galaxy). Black holes are now recognized as important, alongside stars, in bringing about reionization around $z\sim10$ and their long range, radiative influence, which can affect, the subtle business of galaxy formation and evolution may limit the accuracy with which quantitative analyses of large scale structure can be conducted. On the stellar scale, we are about to learn much from general, transient astronomy, now including gravitational radiation, which may transform our understanding of advanced stellar evolution and X-ray binaries.

However, it is with spinning black holes, interpreted as cosmic machines, where most progress seems imminent. It is a very good working hypothesis that they are described by the Kerr metric and that strong field relativity has no surprises in store. LIGO-VIRGO has already affirmed this belief and has the opportunity to challenge it in much more detail in the near future. In order to understand accretion disks and relativistic jets, we have to deal with the classical physics of fluid mechanics, radiative transfer and electromagnetism. The capability of modern simulations is impressive but they are now limited by our ignorance of the principles that control where and how dissipation takes place, especially leading to  the acceleration of non-thermal electrons and positrons. However many of the issues are being forced by observations of rapid $\gamma$-ray variability and imaging polarimetry of the innermost segments of relativistic jets at mm and eventually submm wavelengths.  The growing scope of PIC simulations will also help us to determine the rules that magnetized, relativistic plasma obeys when it dissipates and radiates.  

The major impact of these investigations on LISA will be to give a more quantitative understanding of black hole  demographics and its redshift dependence. We hope to attend this symposium in a decade when LISA is just about to launch!
\ack
We thank Greg Madejski, Jonathan McKinney, Tony Readhead, Sasha Tchekhovskoy, and Craig Walker for advice and collaboration. RA acknowledges the Stanford DARE program and the CA Alliance for Graduate Education and the Professoriate for fellowships.
\section*{References}
\bibliography{biblio.bib}

\begin{thebibliography}{9}
\bibitem{Ackermann2016}Ackermann, M. {\it et al.}2016 ApJ 824 L20
\bibitem{Akiyama2015}Akiyama, K. {\it et al.} 2015 ApJ 807 150
\bibitem{Anantua2016}Anantua, R. 2016 Unpublished thesis, Stanford
\bibitem{Anantua2017}Anantua, R. {\it et al.} 2017 in preparation
\bibitem{Balbus1991}Balbus, S. \& Hawley, J . 1991 ApJ 376 214
\bibitem{Blandford1977}Blandford, R \& Znajek, R. 1977 MNRAS 179 433
\bibitem{Blandford1999}Blandford, R. \& Begelman, M. 1999 MNRAS 303 L1
\bibitem{Blandford2014}Blandford, R. {\it et al.} 2014 Nucl. Phys. B. 256 9
\bibitem{Broderick2016}Broderick, A. {\it et al.} 2016 ApJ 820 137
\bibitem{Gebhardt2011}Gebhardt, K. {\it et al.} 2011 ApJ 729 119
\bibitem{Johnson2015}Johnson, M. {\it et al.} 2015 Science 350 1242
\bibitem{McKinney2012}McKinney, J. {\it et al.} 2012 MNRAS 423 3083
\bibitem{Rathore2005}Rathore, Y. 2005 Unpublished thesis, Caltech 
\bibitem{Siegert2016}Siegert, T. {\it et al.} 2016 Nature 531 341
\bibitem{Soltan1982}Soltan, A. 1982 MNRAS 200 115
\bibitem{Yuan2016}Yuan, Y. 2016 ApJ 828 92
\end{thebibliography}

\end{document}